\documentclass[sigconf,natbib=true,screen=true]{acmart}

\usepackage{acronym}
\usepackage[skip=0pt]{caption}
\usepackage[inline]{enumitem}
\usepackage{booktabs} 
\usepackage{multirow}
\usepackage{array}
\usepackage{pifont}
\usepackage{xspace}
\usepackage{subcaption}
\usepackage{CJKutf8}
\usepackage[para]{footmisc}
\usepackage{color,xcolor}
\usepackage{multirow}
\usepackage{enumitem}
\usepackage{balance}
\usepackage{etoolbox}

\acrodef{QPP4CS}{query performance prediction for conversational search}

\acrodef{CS}{conversational search}
\acrodef{QPP}{query performance prediction}
\acrodef{CIS}{conversational information seeking}
\acrodef{CAsT}{conversational assistance track}
\acrodef{IR}{information retrieval}

\acrodef{MLP}{multilayer perceptron}
\acrodef{MLE}{maximum likelihood estimation}

\theoremstyle{remark}

\AtBeginDocument{%
  \providecommand\BibTeX{{%
    \normalfont B\kern-0.5em{\scshape i\kern-0.25em b}\kern-0.8em\TeX}}}

\newcommand{\CAsT}{CAsT\xspace}
\newcommand{\ORQuAC}{OR-QuAC\xspace}
\newcommand{\QuReTeC}{QuReTeC\xspace}
\newcommand{\QPP}{\ac{QPP}\xspace}
\newcommand{\CS}{\ac{CS}\xspace}

\makeatletter
\@ifpackagelater{acronym}{2020/01/01}
{%
}
{%

\newcommand{\Acf}[1]{\acf{#1}(\textbf{properly capitalized})}
}%
\makeatother

\newcommand{\header}[1]{\vspace*{1mm}\noindent\textbf{#1}.}

\makeatletter
\g@addto@macro\normalsize{%
  \abovedisplayskip 1pt plus1pt 
  \belowdisplayskip 1pt plus1pt
  \abovedisplayshortskip  0pt plus1pt%
  \belowdisplayshortskip  0pt plus1pt
}
\makeatother
\copyrightyear{2023}
\acmYear{2023}
\setcopyright{rightsretained}
\acmConference[SIGIR '23]{Proceedings of the 46th International ACM SIGIR Conference on Research and Development in Information Retrieval}{July 23--27, 2023}{Taipei, Taiwan} \acmBooktitle{Proceedings of the 46th International ACM SIGIR Conference on Research and Development in Information Retrieval (SIGIR '23), July 23--27, 2023, Taipei, Taiwan}\acmDOI{10.1145/3539618.3591919} \acmISBN{978-1-4503-9408-6/23/07}

\makeatletter
\gdef\@copyrightpermission{
  \begin{minipage}{0.3\columnwidth}
   \href{https://creativecommons.org/licenses/by/4.0/}{\includegraphics[width=0.90\textwidth]{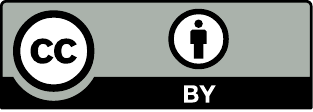}}
  \end{minipage}\hfill
  \begin{minipage}{0.7\columnwidth}
   \href{https://creativecommons.org/licenses/by/4.0/}{This work is licensed under a Creative Commons Attribution International 4.0 License.}
  \end{minipage}
  \vspace{5pt}
}
\makeatother

\author{Chuan Meng}
\orcid{0000-0002-1434-7596}
\affiliation{%
  \institution{University of Amsterdam}
  \city{Amsterdam}
  \country{The Netherlands}
}
\email{c.meng@uva.nl}

\author{Negar Arabzadeh}
\orcid{0000-0002-4411-7089}
\affiliation{%
  \institution{University of Waterloo}
  \city{Waterloo}
  \country{Canada}
}
\email{narabzad@uwaterloo.ca}

\author{Mohammad Aliannejadi}
\orcid{0000-0002-9447-4172}
\affiliation{%
  \institution{University of Amsterdam}
  \city{Amsterdam}
  \country{The Netherlands}
}
\email{m.aliannejadi@uva.nl}

\author{Maarten de Rijke}
\orcid{0000-0002-1086-0202}
\affiliation{%
  \institution{University of Amsterdam}
  \city{Amsterdam}
  \country{The Netherlands}
}
\email{m.derijke@uva.nl}

\begin{document}

\title[Query Performance Prediction: From Ad-hoc to Conversational Search]{Query Performance Prediction: 
\\ From Ad-hoc to Conversational Search}

\renewcommand{\shortauthors}{Chuan Meng, Negar Arabzadeh, Mohammad Aliannejadi, \& Maarten de Rijke}




\begin{abstract}
\Acf{QPP} is a core task in information retrieval.
The \ac{QPP} task is to predict the retrieval quality of a search system for a query without relevance judgments.
Research has shown the effectiveness and usefulness of \QPP for ad-hoc search. 
Recent years have witnessed considerable progress in conversational search (CS).
Effective \QPP could help a \acs{CS} system to decide an appropriate action to be taken at the next turn.
Despite its potential, \QPP for \acs{CS} has been little studied.
We address this research gap by reproducing and studying the effectiveness of existing \QPP methods in the context of \acs{CS}.
While the task of passage retrieval remains the same in the two settings, a user query in \acs{CS} depends on the conversational history, introducing novel \QPP challenges. 
In particular, we seek to explore to what extent findings from \QPP methods for ad-hoc search generalize to three \acs{CS} settings:
\begin{enumerate*}[label=(\roman*)]
\item estimating the retrieval quality of different query rewriting-based retrieval methods,
\item estimating the retrieval quality of a conversational dense retrieval method, and
\item estimating the retrieval quality for top ranks vs.\ deeper-ranked lists.
\end{enumerate*}
Our findings can be summarized as follows:
\begin{enumerate*}[label=(\roman*)]
\item supervised \QPP methods distinctly outperform unsupervised counterparts only when a large-scale training set is available;
\item point-wise supervised \QPP methods outperform their list-wise counterparts in most cases; and 
\item retrieval score-based unsupervised \QPP methods show high effectiveness in assessing the conversational dense retrieval method, ConvDR.
\end{enumerate*} 
\end{abstract}

\maketitle

\acresetall


\vspace*{-2mm}
\section{Introduction}
\Acf{QPP} is an essential task in information retrieval (\acs{IR}\acused{IR}).
It is about estimating the retrieval quality of a search system for a given query without relevance judgments~\citep{he2006query,zhou2007query,zamani2018neural, datta2022deep, datta2022pointwise,ganguly2022analysis}. 
\QPP has been long studied in the \ac{IR} community~\cite{cronen2002predicting}. 
Numerous benefits of \QPP have been identified, including selecting the most effective ranking algorithm for a query~\citep{he2006query,he-2008-using,zamani2018neural} based on the difficulty of the input query.

In \CS there has been significant progress on multiple subtasks~\citep{zamani2022conversational}, including passage retrieval \cite{dalton2020cast,yu2021few}, query rewriting \cite{yu2020few,voskarides2020query}, mixed-initiative interactions \cite{aliannejadi2019asking,zamani2020generating}, response generation~\citep{meng2020refnet,meng2020dukenet,meng2021initiative}, and evaluation~\citep{faggioli2021hierarchical,faggioli2022dependency}.
Specifically, passage retrieval has been the main focus of TREC \CAsT 2019--2022~\citep{dalton2020cast}, where modeling long conversational context for retrieval is shown to be challenging \cite{aliannejadi20harnessing}. 
Moreover, research has shown that mixed-initiative interactions can lead to improved user and system performance~\cite{aliannejadi2019asking,zou2022users}. 
%
As with ad-hoc retrieval, \QPP benefits \CS in multiple ways. For instance, effective \QPP can help a \CS system take appropriate action at the next turn, e.g., take the initiative in asking a clarifying question or saying ``I cannot answer your question'' to the user, instead of giving a low-quality or risky answer when the estimated retrieval quality for the current user query is low~\citep{roitman2019study,arabzadeh2022unsupervised}.

Despite its importance and significance, little research has been done on \QPP for \CS~\citep{meng2023Performance}. 
We take the first steps in this direction by conducting a comprehensive reproducibility study, where we examine a variety of \QPP methods that were originally designed for ad-hoc retrieval in the setting of \CS. 
We aim to characterize the novel challenges of \QPP for \CS and highlight the unique characteristics of this field, while simultaneously assessing the effectiveness of existing \QPP methods in a conversational setting.

In particular, we highlight three main challenges of \QPP applied to \CS that distinguish it from the ad-hoc search setting:
\begin{enumerate}[label=(\arabic*),leftmargin=*,nosep]
\item a user query in a conversation depends on the conversational context, i.e., it may contain omissions, coreferences, or ambiguities, leading to unforeseen \QPP challenges;
\item \QPP for \CS has to predict the performance of novel retrieval approaches, approaches that are specifically designed for \CS;
two main groups of \CS methods have been proposed to solve the query understanding challenge in \CS, i.e., query-rewriting-based retrieval~\citep{mele2020topic,voskarides2020query,yu2020few,vakulenko2021comparison,lin2021multi,wu2022conqrr} and conversational dense retrieval methods~\citep{qu2020open,yu2021few,lin2021contextualized,mao2022curriculum,mao2022curriculum,kim2022saving,mao2022convtrans}. 
\item \QPP for \CS should focus on estimating the retrieval quality for the top-ranked results rather than for a full-ranked list because \CS systems need to return brief responses to adapt to limited-bandwidth interfaces, such as a mobile screen~\citep{zamani2022conversational}.
\end{enumerate}
In this reproducibility paper, we design our experiments inspired by these \CS characteristics and examine whether established findings on \QPP for ad-hoc search still hold under these conditions.
Specifically, we study the following findings from the literature on \QPP for ad-hoc search:
\begin{enumerate*}[label=(\roman*)]
\item supervised \QPP methods outperform unsupervised \QPP methods~\citep{zamani2018neural,hashemi2019performance,arabzadeh2021bert,datta2022deep,chen2022groupwise,datta2022pointwise};
\item list-wise supervised \QPP methods outperform their point-wise counterparts~\citep{chen2022groupwise,datta2022pointwise}; and
\item retrieval score-based unsupervised \QPP methods perform poorly in estimating the retrieval quality of neural-based retrievers~\citep{datta2022relative,hashemi2019performance}.
\end{enumerate*} 
By examining each of these \QPP-for-ad-hoc-search findings listed above in the setting of \CS, we aim to characterize the problem of \QPP applied to \CS, with novel findings and directions for future research as additional outcomes. 

In this paper, we conduct experiments on three \CS datasets:
\begin{enumerate*}[label=(\roman*)]
\item \CAsT-19~\citep{dalton2020cast}, 
\item \CAsT-20~\citep{Dalton2020CAsT2T}, and \item \ORQuAC~\citep{qu2020open}.
\end{enumerate*}
Our experiments show that, in the setting of \CS,
\begin{enumerate*}[label=(\roman*)]
\item supervised \QPP methods distinctly outperform unsupervised counterparts only when a large amount of training data is available; unsupervised \QPP methods show strong performance in a few-shot setting and when predicting the retrieval quality for deeper ranked lists;
\item point-wise supervised \QPP methods outperform their list-wise counterparts in most cases; however, list-wise \ac{QPP} methods show a slight advantage in a few-shot setting and when predicting the retrieval quality for deeper ranked lists; and 
\item retrieval score-based unsupervised \QPP methods show high effectiveness in estimating the retrieval quality of a conversational dense retrieval method, ConvDR, either for top ranks or deeper ranked lists.
\end{enumerate*}

\vspace*{-1mm}
\section{Preliminaries and task definition}
We recap the definition of the \QPP task in the context of ad-hoc search.
Generally, given a query $q$, a collection of documents $D$, an ad-hoc retrieval method $M$ and the ranked list with top-$k$ ranked documents \smash{$D_{q;M}^{k}=[d_1, d_2, \dots,d_k]$} returned by the retriever $M$ over the collection $D$ with respect to the query $q$, a \QPP method $f$ estimates the retrieval quality of the ranked list \smash{$D_{q;M}^{k}$} with respect to the query $q$, formally:
\begin{equation}
\phi=f(q,D_{q;M}^{k},D)\in \mathbb{R}~,
\label{define:qpp}
\end{equation}
where $\phi$ indicates the retrieval quality of the ad-hoc retriever $M$ in response to the query $q$; the retrieval quality $\phi$ can depend on collection-based statistics.

Next, we define the task of \QPP for \CS.
The \CS task is to find relevant items for each query in a multi-turn conversation $Q=\{q_t\}^n_{t=1}$~\citep{dalton2020cast}, where $n$ is the number of turns in a conversation. 
Unlike traditional ad-hoc search, the query $q_t$ at turn $t$ may contain omissions, coreferences, or ambiguities, making it hard for ad-hoc search methods to capture the underlying information need of the query $q_t$~\citep{yu2021few}.
Two main groups of \CS methods have been proposed to solve the query understanding challenge in \CS, i.e., query rewriting-based retrieval~\citep{mele2020topic,voskarides2020query,yu2020few,vakulenko2021comparison,lin2021multi} and conversational dense retrieval methods~\citep{yu2021few,lin2021contextualized,mao2022curriculum}.
Query rewriting-based retrieval methods first rewrite the query $q_t$ into a self-contained query $q'_t$ with the conversational history $Q_{1:t-1}=q_1,q_2,\dots,q_{t-1}$, and then reuse ad-hoc search methods using the rewritten query $q'_t$ as input.
When estimating the retrieval quality of this group of \CS methods, we define \QPP for \CS as:
\begin{equation}
\begin{split}
\phi_t=f(q'_t,D_{q'_t;M}^{k},D)\in \mathbb{R}~,
\end{split}
\label{define:qpp4cs1}
\end{equation}
where, given the query rewrite $q'_t$, the ranked list of documents \smash{$D_{q'_t;M}^{k}$} retrieved by an ad-hoc search method $M$ for the query rewrite $q'_t$, predicts $\phi_t$ that is indicative of the retrieval quality of the method in response to the rewritten query $q'_t$.

Conversational dense retrieval methods train a query encoder to encode the current query $q_t$ and the conversation history $Q_{1:t-1}$ into a contextualized query embedding that is used to represent the information need of the current query in a latent space~\citep{yu2021few,mao2022curriculum}.
However, existing \QPP methods do not have such a special module to understand the noisy raw utterances $Q_{1:t}$; directly feeding the raw utterances $Q_{1:t}$ into \QPP methods may fuse them.
Thus, when estimating the retrieval quality of a conversational dense retrieval method, we still feed a query rewrite $q'_t$ instead of the raw utterances $Q_{1:t}$ into \QPP methods, formally:
\begin{equation}
\begin{split}
\phi_t=f(q'_t,D_{Q_{1:t};M}^{k},D)\in \mathbb{R}~,
\end{split}
\label{define:qpp4cs2}
\end{equation}
where \smash{$D_{Q_{1:t};M}^{k}$} is the ranked list retrieved by a conversational dense retrieval method $M$ in response to the raw utterances $Q_{1:t}$.
\vspace*{-1mm}
\section{Reproducibility Methodology}

We describe our research questions and the experiments designed to address them. We also describe our experimental setup.

\vspace*{-1mm}
\subsection{Research questions}
We address the following research questions:
\begin{enumerate}[label=(\textbf{RQ\arabic*})]
\item Does the performance of \QPP methods for ad-hoc search generalize to \CS when estimating the retrieval quality of different query rewriting-based retrieval methods? \label{RQ1}

\item  Does the performance of \QPP methods for ad-hoc search generalize to \CS when estimating the retrieval quality of a conversational dense retrieval method? Is the \QPP effectiveness influenced by the choice of query rewrites?\label{RQ2}

\item  What is the performance difference between \QPP methods when predicting the retrieval quality for top-ranked items vs.\ for longer-ranked lists?\label{RQ3}
\end{enumerate}

\vspace*{-1mm}
\subsection{Experimental design}
Next, we describe the experiments aimed at answering our research questions. 
Our main goal is to study the reproducibility of ad-hoc \QPP methods in the \CS setting.
We compare the performance of unsupervised and supervised \QPP methods on three \CS datasets.
Specifically, we conduct the following experiments:
\begin{enumerate}[label=\textbf{E\arabic*}, leftmargin=*, nosep]
    \item To address \ref{RQ1}, we estimate the retrieval quality of BM25 with three query rewriting methods, namely, T5, QuReTeC, and perfect rewriting (human-rewritten)~\cite{dalton2020cast}.
    Note that \QPP methods and BM25 always share the same query rewrites.\label{E1}

    \item To address \ref{RQ2}, we study the performance of \QPP methods for a conversational dense retrieval method, ConvDR~\citep{yu2021few}, on all three datasets. As ConvDR directly models the raw conversation context, no query rewriting step is required. However, no existing \QPP methods can model raw conversations. Hence, we study the effect of feeding different query rewrites into \QPP methods when predicting the performance of Conv\-DR.\label{E2}
    
    \item To address \ref{RQ3}, we apply the \QPP methods on evaluation metrics at different depths. We utilize nDCG@3 and nDCG@100 and analyze how \QPP performance is affected by the ranking depth. We also consider Recall@100 to study the effectiveness of \QPP for first-stage \CS rankers, where high recall is desired.\label{E3}
\end{enumerate}

\vspace*{-1mm}
\subsection{Experimental setup}
\label{sec:setup}

\textbf{\QPP methods.}
We analyze a variety of unsupervised/supervised \QPP methods.
For unsupervised ones, we consider clarity-based and score-based methods because they have been widely used in the literature.
We consider more score-based ones since they have shown great effectiveness~\citep{carmel2010estimating}.
We consider one clarity-based method:
\begin{itemize}[leftmargin=*,nosep]
\item  Clarity~\cite{cronen2002predicting} quantifies the degree of ambiguity of a query w.r.t.\ a collection of documents. Specifically, it measures the KL divergence between a relevance model~\citep{lavrenko2001relevance} induced from top-ranked documents and a language model induced from the collection:
\begin{equation}
\mathit{Clarity}(q,D_{q;M}^{k},D)=\sum_{w \in V} P(w|D_{q;M}^{k})\log \frac{P(w|D_{q;M}^{k})}{P(w|D)},
\end{equation}
where $w$ and $V$ denote a term and the entire vocabulary of the collection, respectively. 
The conjecture is that the larger the KL divergence is, the better the retrieval quality is.

\end{itemize}


\noindent We consider five score-based \QPP methods:
\begin{itemize}[leftmargin=*,nosep]
\item  Weighted information gain (WIG)~\cite{zhou2007query} measures the divergence of retrieval scores of top-ranked documents from those of the entire corpus: the higher the divergence is, the better the retrieval quality is~\citep{shtok2012predicting,tao2014query,zamani2018neural}.
WIG is formulated as:
\begin{equation}
\mathit{WIG}(q,D_{q;M}^{k},D)=\frac{1}{k}\sum_{d \in  D_{q;M}^{k}}\frac{1}{\sqrt{|q|}} (\mathit{Score}(q;d)-\mathit{Score}(q;D)),
\end{equation}
where $\mathit{Score}(q;d)$ and $\mathit{Score}(q;D)$ are the retrieval scores of document $d$ and the entire collection $D$, respectively; $|q|$ is $q$'s length.
\item Normalized query commitment (NQC)~\cite{shtok2012predicting} measures the standard deviation of retrieval scores of top-ranked documents; the standard deviation is normalized by the retrieval score of the entire collection $D$.
The higher the standard deviation is, the better the retrieval quality is assumed to be.
NQC is modeled as: 
\begin{equation}
\mbox{}\hspace*{-2mm}
NQC(q,D_{q;M}^{k},D)=\frac{1}{\mathit{Score}(q;D)} \sqrt{\frac{1}{k}\sum_{d \in  D_{q;M}^{k}}\!\!(\mathit{Score}(q;d)-\mu)^2},
\hspace*{-3mm}\mbox{}
\end{equation}
where $\mu$ is the mean retrieval score of the top-ranked documents.
\item $\sigma_{max}$~\citep{perez2010standard} is based on the standard deviation of retrieval scores of ranked documents but finds the most suitable ranked list size $k$ for each query. The intuition is that most of the retrieved documents in a ranked list obtain a low retrieval score; considering such non-relevant documents would hurt \QPP effectiveness. $\sigma_{max}$ computes the standard deviation at each point in the ranked list and selects the maximum standard deviation so as to reduce the impact of the documents with a low retrieval score.
\item  n($\sigma_{x\%}$)~\citep{cummins2011improved}, similar to $\sigma_{max}$, also uses a dynamic number of documents to calculate the standard deviation for each query, but only considers the documents whose retrieval scores are at least $x\%$ of the top retrieval score. The calculated standard deviation is normalized by query length.

\item  Score magnitude and variance (SMV)~\cite{tao2014query} argues that WIG and NQC mainly consider the magnitude and the variance of retrieval scores, respectively. SMV takes both aspects into consideration:
\begin{equation}
    SMV(q,D_{q;M}^{k},D)=\frac{\frac{1}{k}\sum_{d \in D_{q;M}^{k}}(\mathit{Score}(q;d)\lvert\ln\frac{\mathit{Score}(q;d)}{\mu}\rvert)}{\mathit{Score}(q;D)}, 
\end{equation}
where $\mathit{Score}(q;d)$ denotes score magnitude while $\lvert\ln\frac{\mathit{Score}(q;d)}{\mu}\rvert)$ represents score variance.
\end{itemize}  


\noindent Recent studies show that BERT-based supervised \QPP methods~\citep{hashemi2019performance,arabzadeh2021bert,datta2022pointwise,chen2022groupwise} outperform other neural-based supervised \QPP methods, such as NeuralQPP~\citep{zamani2018neural} and Deep-QPP~\citep{datta2022deep}.
Thus, we consider three competitive BERT-based supervised \QPP methods: 
    \begin{itemize}[leftmargin=*,nosep]
    \item  NQA-QPP~\cite{hashemi2019performance} is the first supervised \QPP method based on BERT. It feeds the standard deviation of retrieval scores, BERT representations for the given query and query-document pairs into a feed-forward neural network for estimating the retrieval quality. 
    \item  BERT-QPP~\cite{arabzadeh2021bert} feeds the given query and the top-ranked document into BERT, followed by a linear layer for estimating the retrieval quality. We use the cross-encoder version of BERT-QPP as it outperforms the bi-encoder version.
    \item  qppBERT-PL~\cite{datta2022pointwise} is a listwise-document method. It splits the top-ranked documents into chunks and then uses BERT to encode all query-document pairs in each chunk; a sequence of query-document BERT representations in a chunk is fed into an LSTM and linear layers to predict the number of relevant documents in the chunk. A weighted average of the number of relevant documents across all chunks is calculated as the retrieval quality.
\end{itemize}

\noindent%
We do not include BERT-groupwise-QPP~\citep{chen2022groupwise}. 
It is another list-wise supervised \QPP method, which uses cross-query information but it cannot be directly applied in a \CS setting, as it would access the future next turn query $q_{t+1}$ when estimating the difficulty of the current query $q_t$ during inference, which is unrealistic in \CS. 

\header{Query rewriting methods}
We adopt the following query rewriting techniques/data in the passage retrieval and \QPP process:
\begin{enumerate*}[label=(\roman*)]
    \item T5 rewriter\footnote{\url{https://huggingface.co/castorini/t5-base-canard}} is fine-tuned on CANARD~\cite{elgohary2019can} query rewriting dataset;
    \item \QuReTeC~\cite{voskarides2020query} is a BERT-based term expansion query rewriting method. We use the checkpoint released by the author;\footnote{\url{https://github.com/nickvosk/sigir2020-query-resolution}} and
    \item Human is the human-generated oracle query rewriting model obtained from the ground-truth data annotations.
\end{enumerate*}

\header{\CS methods to be evaluated for retrieval quality}
We estimate the retrieval quality of two groups of \CS methods: query rewriting-based retrieval and conversational dense retrieval methods.
For the former, we consider:
\begin{enumerate*}[label=(\roman*)]
    \item T5+BM25 rewrites queries using the T5 rewriter and ranks documents using BM25\footnote{We use Pyserini BM25 with the default parameters k1=0.9, b=0.4.};
    \item \QuReTeC+BM25~\cite{voskarides2020query} performs query resolution using \QuReTeC, followed by BM25 retrieval; and
    \item Human+BM25 uses the ground-truth query rewrites to rank documents using BM25.
\end{enumerate*}
For the latter, we consider ConvDR~\cite{yu2021few} and use the code released by the author.\footnote{\url{https://github.com/thunlp/ConvDR}}
All \CS methods return the top-1000 documents per query.


\header{Datasets}
We consider three \CS datasets:
\begin{enumerate*}[label=(\roman*)]
    \item \CAsT-19~\citep{dalton2020cast} is constructed manually
to mimic a realistic conversation on a specific topic; in this dataset, a later query turn often depends on its previous queries;
    \item \CAsT-20~\citep{Dalton2020CAsT2T} is more realistic and
complex because the information needs of queries are derived from
commercial search logs and queries can refer to previous system responses; and
    \item \ORQuAC~\citep{qu2020open} is a large-scale synthetic \ac{CS} dataset built on a conversational QA dataset, QuAC~\cite{choi2018quac}; there is usually only one annotated relevant item for each query in this dataset.
\end{enumerate*}
All three datasets provide self-contained queries rewritten by humans for all raw queries.
Table~\ref{table:data} lists details of the datasets.

\header{Evaluation}
A common method for evaluating \QPP performance is to assess the correlation between the actual and predicted performance of a query set. 
Pearson's $\rho$, Kendall's $\tau$, and Spearman's $\rho$ correlation coefficients are widely used.
We report the correlation based on the major metrics adopted by TREC CAsT~\citep{dalton2020cast}, namely, nDCG@3 for high ranks and nDCG@100 for deeper ranked lists. 
As mentioned above, we also adopt Recall@100 to investigate the performance of \QPP when evaluating first-stage \CS retrievers.

\header{Implementation details}
We implement all \QPP methods using Pytorch.\footnote{\url{https://pytorch.org/}}
For unsupervised \QPP methods, we use hyperparameters that have been shown to be effective by previous studies.
Following~\citep{zhou2007query}, $k$ is set to 5 for WIG.
As suggested by~\citep{shtok2012predicting,tao2014query}, $k$ is set to 100 for NQC and SMV; following~\citep{tao2014query}, we use the average retrieval score of the top-1000 documents as the corpus score $\mathit{Score}(q;D)$.
Following~\citep{cummins2011improved}, we set $x$ to 50 for n($\sigma_{x\%}$).
$\sigma_{max}$ does not use any hyperparameters.
Following~\citep{shtok2012predicting}, we use the Clarity variant that uses the sum-normalized retrieval scores (from BM25 or ConvDR in our setting) for weighing documents when constructing a relevance model~\citep{lavrenko2001relevance}; our preliminary experiments showed that this variant performed better than the original Clarity that uses query-likelihood scores to weight documents; we induce the relevance model using the top 100 documents and clip the relevance model at the top-100 terms cutoff~\citep{shtok2010using}.


For all supervised \QPP methods, we use bert-base-uncased,\footnote{\url{https://github.com/huggingface/transformers}} a fixed learning rate (0.00002), and the Adam optimizer~\citep{kingma2014adam}.
All methods are trained and inferred on an NVIDIA RTX A6000 GPU.
Following \citep{yu2021few,mao2022curriculum}, all training on \CAsT-19 or \CAsT-20 uses five-fold cross-validation; we use the data split from \citep{yu2021few} and train all supervised \QPP methods for 5 epochs. 
For training on \ORQuAC, we train all \QPP methods for 1 epoch on the training set of \ORQuAC; we feed \QPP methods with human-rewritten queries and train them to estimate the retrieval quality of BM25 with human-rewritten queries.
To address the data scarcity on \CAsT-19 and \CAsT-20, we consider a \textit{warm-up} setting where we first pre-train supervised \QPP methods on the training set of \ORQuAC for one epoch, followed by the five-fold cross-validation training for 5 epochs on \CAsT.
For future reproducibility, our code and data resources are available at \url{https://github.com/ChuanMeng/QPP4CS}.

\begin{table}[!t]
\centering
\caption{Actual retrieval quality of the \CS methods used in this paper in terms of nDCG@3.}
\label{table:actual}
\setlength{\tabcolsep}{0.8mm}
\resizebox{\linewidth}{!}{%
\begin{tabular}{l@{}c c c}
\toprule
\multirow{1}{*}{} & \multicolumn{1}{c}{CAsT-19}  & \multicolumn{1}{c}{CAsT-20}  & \multicolumn{1}{c}{OR-QuAC}  \\ 
\midrule
 T5-based query rewriter  + BM25  & 0.330   & 0.170    & 0.218 \\
 QuReTeC-based query rewriter + BM25 & 0.338  & 0.172   & 0.249\\
 Human query rewriter + BM25  & 0.360   & 0.257    & 0.309\\
 \midrule
 ConvDR  & 0.471   & 0.343   & 0.614  \\
\bottomrule
\end{tabular}
}
\end{table}

\begin{table}[!t]
\centering
\caption{Data statistics of CAsT-19, CAsT-20 and OR-QuAC.}
\label{table:data}
\setlength{\tabcolsep}{0.8mm}
\resizebox{\linewidth}{!}{%
\begin{tabular}{l@{}rrrrr}
\toprule
\multirow{1}{*}{} & \multicolumn{1}{c}{CAsT-19}  & \multicolumn{1}{c}{CAsT-20}  & \multicolumn{3}{c}{OR-QuAC}  \\ 
\cmidrule(lr){2-2} \cmidrule(lr){3-3}  \cmidrule(lr){4-6}
 & test & test  & train & valid  & test \\ 
\midrule
\#conversations  &50 & 25 & 4,383 & 490 & 771 \\
\#conversations (judged) & 20 & 25 & -- & -- & -- \\
\#questions  & 479    & 216  & 31,526 & 3,430 & 5,571   \\
\#questions (judged) & 173   & 208 & --  & -- & -- \\
\#documents   &\multicolumn{2}{c}{38M}   &\multicolumn{3}{c}{11M}   \\
\bottomrule
\end{tabular}
}
\end{table}

\section{Results and discussions}
\label{res}
Our experiments revolve around three main findings from the literature on \QPP for ad-hoc search:
\begin{enumerate*}[label=(\roman*)]
\item supervised \QPP methods outperform unsupervised \QPP methods~\citep{zamani2018neural,hashemi2019performance,arabzadeh2021bert,datta2022deep,chen2022groupwise,datta2022pointwise};
\item list-wise supervised \QPP methods outperform their point-wise counterparts~\citep{chen2022groupwise,datta2022pointwise}; and
\item retrieval score-based unsupervised \QPP methods perform poorly in estimating the retrieval quality of neural-based retrievers~\citep{datta2022relative,hashemi2019performance}.
\end{enumerate*}
We study whether the findings listed above continue to hold for \QPP methods in \CS.

\vspace*{-2mm}
\subsection{Assessing query rewriting-based retrieval}
\label{res:rq1}

\subsubsection{Overall performance}
\label{res:rq1:overall}

To answer \ref{RQ1}, we examine the results of Experiment~\ref{E1}, where we run \QPP methods estimating the retrieval quality of BM25 with three query rewriting methods~(T5+BM25, QuReTeC+BM25, and Human+BM25). 
For all supervised \QPP methods on \CAsT, we further consider their variants that are first pre-trained on the training set of \ORQuAC for one epoch before five-fold cross-validation training on \CAsT.
See Table~\ref{tab:RQ1}.
Note that \QPP methods and BM25 always share the same query rewrites.
Overall, feeding T5/QuReTeC query rewrites into \QPP methods to estimate the retrieval quality of BM25 is effective, compared to the case of feeding perfect self-contained queries rewritten by humans.
We have two specific observations.

\begin{table*}[]
\caption{
Outcomes of Experiment~\ref{E1}.
Performance of \QPP methods on three \CS datasets: Pearson's $r$, Kendall's $\tau$, and Spearman's $\rho$ correlation coefficients with nDCG@3, for estimating the retrieval quality of three query rewriting-based retrieval methods (BM25 fed with T5-based, QuReTeC-based, and human-written query rewrites).
\textit{Warm-up} means the \QPP method is first pre-trained on the training set of \ORQuAC for one epoch.
All coefficients are statistically significant (t-test, $p < 0.05$) except the ones in \emph{italics}.
The best value in each column is marked in \textbf{bold}, and the second best is \underline{underlined}.
}
\label{tab:RQ1}
\begin{tabular}{ll rrr rrr rrr}
\toprule
 & \multicolumn{1}{c}{} & \multicolumn{3}{c}{T5+BM25} & \multicolumn{3}{c}{QuReTeC+BM25} & \multicolumn{3}{c}{Human+BM25} \\
 \cmidrule(lr){3-5}  \cmidrule(lr){6-8}  \cmidrule(lr){9-11}
Datasets & \multicolumn{1}{l}{QPP methods} & \multicolumn{1}{c}{P-$\rho$} & \multicolumn{1}{c}{K-$\tau$} & \multicolumn{1}{c}{S-$\rho$} & \multicolumn{1}{c}{P-$\rho$} & \multicolumn{1}{c}{K-$\tau$} & \multicolumn{1}{c}{S-$\rho$} & \multicolumn{1}{c}{P-$\rho$} & \multicolumn{1}{c}{K-$\tau$} & \multicolumn{1}{c}{S-$\rho$} \\
\midrule
\multirow{12}{*}{CAsT-19} & Clarity & 0.321  & 0.234  & 0.330  & 0.327  & 0.211 & 0.304 & 0.359 & 0.231  & 0.335  \\
 & WIG & 0.436 & 0.232 & 0.452 & 0.354 &  0.250  & 0.356  & 0.409 & 0.293 & 0.414 \\
 & NQC & 0.348 & 0.246  & 0.354 & 0.286 & 0.190 &  0.275 & 0.334 & 0.234 & 0.335 \\
 & $\sigma_{max}$ & 0.442 & \underline{0.354} & 0.501 & 0.351 & 0.251 & 0.357 & \underline{0.410} & \textbf{0.312}  & \textbf{0.441} \\
 & n($\sigma_{x\%}$)  &  0.430 &  0.332 & 0.466 & 0.348 & 0.259 & 0.364  & 0.407 & \underline{0.307}   & \underline{0.430} \\
 & SMV & 0.344 & 0.250 & 0.360 & 0.289 & 0.188 & 0.273 & 0.326 & 0.230 & 0.333\\
& NQA-QPP & 0.188 & \textit{0.047} & \textit{0.072}     & \textit{-0.016} & \textit{0.010} & \textit{0.014}     & 0.152 & \textit{0.069} & \textit{0.099}     \\
& BERTQPP & 0.440 & 0.307 & 0.424     & 0.352 & 0.272 & 0.395     & 0.270 & 0.188 & 0.271     \\
& qppBERT-PL & 0.414 & 0.296 & 0.421     & \underline{0.392} & \underline{0.298} & \underline{0.406}     & 0.292 & 0.196 & 0.280     \\
& NQA-QPP (warm-up) & \textbf{0.538} & \textbf{0.357} & \textbf{0.510}     & \textbf{0.420} & \textbf{0.301} & \textbf{0.428}     & 0.331 & 0.230 & 0.336     \\
& BERTQPP (warm-up) & \underline{0.526} & \textbf{0.357} & \underline{0.503}     & 0.369 & 0.264 & 0.384     & \textbf{0.418} & 0.282 & 0.411     \\
& qppBERT-PL (warm-up) & 0.317 & 0.218 & 0.313     & 0.330 & 0.232 & 0.326     & 0.297 & 0.190 & 0.277     \\
  \midrule
\multirow{12}{*}{CAsT-20} & Clarity & \underline{0.258} & 0.191  & 0.259  & \textit{0.099}  & \textit{0.061} & \textit{0.085}  & \textit{0.127}  & \textit{0.089}  & \textit{0.121} \\
 & WIG & 0.248 & \textbf{0.251} & \textbf{0.339} & 0.245 & 0.163 & 0.222  & \underline{0.307} & 0.222 & 0.317\\
 & NQC & 0.150 & \underline{0.235} &  \underline{0.316} &  0.198 &  0.189 & 0.259  & 0.286 & \textbf{0.266} & \textbf{0.370} \\
 & $\sigma_{max}$ & 0.179 & 0.221 & 0.304 & 0.207 & 0.168 &  0.230 &  0.241 &  0.199 & 0.283\\
 & n($\sigma_{x\%}$)  & 0.178 & 0.225 & 0.304 &  0.182 & 0.133 & 0.181 & 0.213 & 0.167 & 0.237 \\
 & SMV &  0.139 &  0.219 & 0.298 & 0.189 & 0.163 & 0.227  & 0.264 & \underline{0.260} & \underline{0.363} \\
& NQA-QPP & \textit{0.001} & \textit{0.067} & \textit{0.093}     & \textit{-0.064} & \textit{-0.082} & \textit{-0.111}     & \textit{0.086} & \textit{-0.011} & \textit{-0.012}     \\
& BERTQPP & \textit{0.042} & \textit{-0.009} & \textit{-0.007}     & 0.172 & 0.145 & 0.196     & 0.194 & 0.110 & 0.159     \\
& qppBERT-PL & \textit{0.131} & 0.125 & 0.159     & 0.175 & 0.150 & 0.185     & \textit{0.043} & \textit{0.015} & \textit{0.021}     \\
& NQA-QPP (warm-up) & \textbf{0.274} & 0.170 & 0.227     & 0.190 & 0.149 & 0.201     & 0.231 & 0.155 & 0.222     \\
& BERTQPP (warm-up) & 0.207 & 0.171 & 0.236     & \textbf{0.403} & \textbf{0.301} & \textbf{0.409}     & \textbf{0.336} & 0.227 & 0.318     \\
& qppBERT-PL (warm-up) & 0.228 & 0.213 & 0.275     & \underline{0.317} & \underline{0.268} & \underline{0.335}     & \textit{0.094} & \textit{0.095} & \textit{0.130}     \\
  \midrule
\multirow{9}{*}{OR-QuAC} & Clarity & 0.090 & 0.085  & 0.110 & 0.110 & 0.103 & 0.133 & 0.076  & 0.069  & 0.091 \\
 & WIG & 0.247 & 0.235 & 0.304 & 0.290 & 0.270  & 0.350 & 0.257  & 0.241  & 0.316 \\
 & NQC & 0.251 & 0.274 & 0.355 & 0.290 & 0.311 & 0.404 & 0.276 & 0.291 & 0.381 \\
 & $\sigma_{max}$ & 0.317 & 0.279 & 0.359 & 0.367 & 0.316 &  0.406 &  0.412&  0.367 & 0.474\\
 & n($\sigma_{x\%}$)  & 0.181 & 0.172 & 0.223 & 0.229 & 0.209 & 0.270 & 0.245  & 0.193   & 0.252 \\
 & SMV & 0.204 & 0.239 & 0.310 & 0.239  & 0.273  & 0.355 & 0.194 & 0.232 & 0.304\\
& NQA-QPP & \textbf{0.781} & \textbf{0.566} & \textbf{0.695}     & \textbf{0.792} & \textbf{0.591} & \textbf{0.725}     & \textbf{0.809} & \textbf{0.621} & \textbf{0.767}     \\
& BERTQPP & \underline{0.678} & 0.434 & 0.546     & \underline{0.692} & 0.476 & \underline{0.598}     & \underline{0.725} & \underline{0.527} & \underline{0.666}     \\
& qppBERT-PL & 0.594 & \underline{0.507} & \underline{0.576}     & 0.617 & \underline{0.526} & 0.597     & 0.618 & 0.525 & 0.600     \\
  \bottomrule
\end{tabular}
\end{table*}

First, when applied to \CS, supervised \QPP methods only have a distinct advantage over their unsupervised counterparts when training data is sufficient.
Specifically, on \ORQuAC, where training data is ample, all supervised \QPP methods perform better than unsupervised methods when assessing BM25 with all three query rewriters.
NQA-QPP achieves state-of-art performance on \ORQuAC.
On \CAsT-19, the performance of unsupervised \QPP methods is comparable to the performance of supervised ones only using five-fold cross-validation. 
However, on \CAsT-20, where the information needs of queries are derived from
commercial search logs and so query understanding is much harder than \CAsT-19, unsupervised \QPP methods perform better than their supervised counterparts only using five-fold cross-validation.
Warming up on the training set of \ORQuAC brings about improvement in supervised \QPP methods in most cases.
On \CAsT-19, NQA-QPP with warm-up performs better than all unsupervised methods given T5/QuReTeC query rewrites.
Nevertheless, on \CAsT-20, even after warming up, supervised methods do not have a distinct advantage.
We think it is because all supervised \QPP methods need to be fed with queries and the difficulty of query understanding on \CAsT-20 limits their performance.
Conversely, the prediction of score-based unsupervised methods does not depend on the input queries, reducing the impact of query understanding.
The performance of qppBERT-PL drops after warming up on \ORQuAC in most cases.
We speculate that this is due to the distribution shift between \CAsT and \ORQuAC: qppBERT-PL predicts the number of relevant documents in each chunk of a ranked list, and the number of relevant documents for each query in \CAsT is significantly larger than in \ORQuAC. 
Therefore, after warming up, qppBERT-PL's prediction of the relevant document count is biased towards the number of relevant documents in \ORQuAC.


Second, in most cases, point-wise supervised \QPP methods such as NQA-QPP and BERTQPP outperform the list-wise supervised method qppBERT-PL.
Without considering warming up, qppBERT-PL has a slight advantage over its point-wise counterparts.
E.g., qppBERT-PL achieves a better performance in predicting the performance of QuReTeC+BM25, Human+BM25 on \CAsT-19, and T5+BM25, QuReTeC+BM25 on \CAsT-20.
qppBERT-PL's list-wise training scheme learns from interactions between a query and all documents in a ranked list, providing the model with more training signals and better use of limited training data. 

%

   
\begin{figure*}[!t]
  \centering
  \includegraphics[clip, trim=0cm 5cm 0.5cm 3.85cm,scale=0.7]{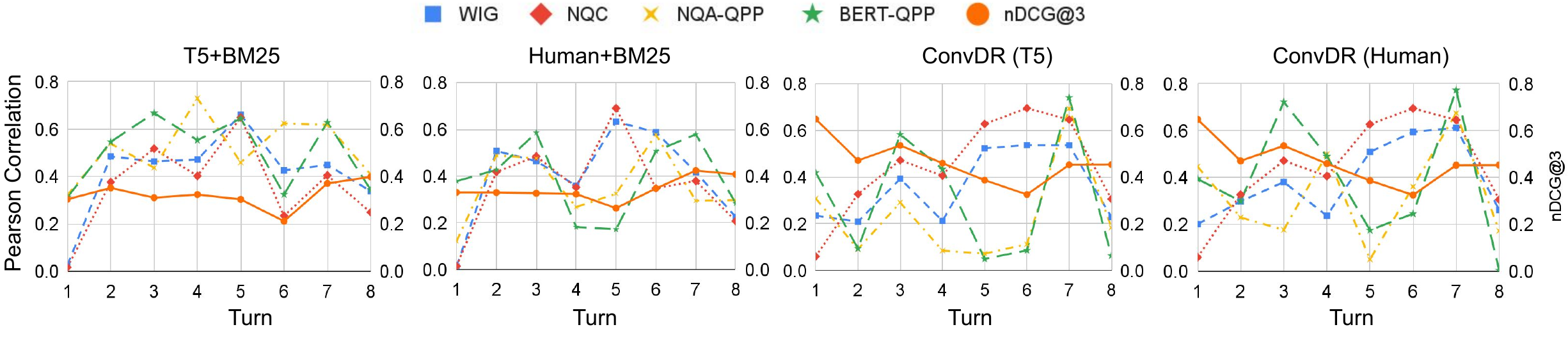}
  \caption{\QPP effectiveness on each turn of conversations in CAsT-19.
  Pearson's $r$ correlation between the actual nDCG@3 scores of the queries with the same turn number and their estimated retrieval quality is calculated per turn.
}
    \label{fig:turn}
\end{figure*}

\subsubsection{Turn-wise \QPP effectiveness}
\label{turnbm25}
We study the \QPP effectiveness on each turn of conversation on \CAsT-19; we report the turn-wise effectiveness of 2 unsupervised (WIG, NQC) and 2 supervised methods (NQA-QPP with warm-up, BERT-QPP with warm-up) when they assess BM25 with T5-based and human-written query rewrites.
The results are presented in the two leftmost subfigures in Figure~\ref{fig:turn}.
We also introduce the turn-wise actual retrieval quality in terms of nDCG@3 in each subfigure.
%
As illustrated in both subfigures, all \QPP methods exhibit lower performance at the first turn and at the deeper turn 8.
There is a correlation between actual retrieval quality and \QPP effectiveness: BERT-QPP effectiveness always drops as the actual retrieval quality drops;
in contrast, in the case of T5+BM25, NQA-QPP performs better as the actual retrieval quality drops at turn 6; in the case of human+BM25, WIG and NQC show better performance as the actual retrieval quality drops at turn 5.





\subsection{Assessing conversational dense retrieval}

\subsubsection{Overall performance}
To answer \ref{RQ2}, we examine the results of \ref{E2}.
We apply \QPP methods fed with three types of query rewrites to estimate the retrieval quality of the conversational dense retrieval method ConvDR.
See Table~\ref{tab:RQ2}. 
Note that the results of NQC, $\sigma_{max}$ and SMV are invariant to different types of query rewrites because they only depend on retrieval scores;
Clarity is also invariant to query rewrites because we use the Clarity variant from \citep{shtok2012predicting}; see Section~\ref{sec:setup} for more information about implementation details. 
We have four main observations.

\begin{table*}[]
\caption{
Outcomes of Experiment~\ref{E2}.
Performance of \QPP methods on three \CS datasets: Pearson's $r$, Kendall's $\tau$, and Spearman's $\rho$ correlation coefficients with nDCG@3, for estimating the retrieval quality of ConvDR (fed with T5-based, QuReTeC-based, and human-written query rewrites).
All coefficients are statistically significant (t-test, $p < 0.05$) except the ones in \emph{italics}.
The best value in each column is marked in \textbf{bold}, and the second best is \underline{underlined}.
}
\label{tab:RQ2}
\begin{tabular}{ll rrr rrr rrr}
\toprule
 & \multicolumn{1}{c}{} & \multicolumn{3}{c}{T5} & \multicolumn{3}{c}{QuReTeC} & \multicolumn{3}{c}{Human} \\
 \cmidrule(lr){3-5}  \cmidrule(lr){6-8}  \cmidrule(lr){9-11}
 Datasets & \multicolumn{1}{l}{{QPP methods}} & \multicolumn{1}{c}{P-$\rho$} & \multicolumn{1}{c}{K-$\tau$} & \multicolumn{1}{c}{S-$\rho$} & \multicolumn{1}{c}{P-$\rho$} & \multicolumn{1}{c}{K-$\tau$} & \multicolumn{1}{c}{S-$\rho$} & \multicolumn{1}{c}{P-$\rho$} & \multicolumn{1}{c}{K-$\tau$} & \multicolumn{1}{c}{S-$\rho$} \\
 \midrule
\multirow{12}{*}{CAsT-19} & Clarity & 0.257  & 0.176  & 0.257 & 0.257 & 0.176 & 0.257  & 0.257 & 0.176  & 0.257  \\
 & WIG & \underline{0.387} & 0.274  & 0.395  & \underline{0.388}  & 0.266 & 0.381  & \underline{0.412}  & \underline{0.285}  & \underline{0.408} \\
 & NQC & \textbf{0.431} & \textbf{0.307}  & \textbf{0.438}  & \textbf{0.431} & \textbf{0.307}  & \textbf{0.438}  & \textbf{0.431} & \textbf{0.307}  & \textbf{0.438} \\
 & $\sigma_{max}$ & 0.378 & 0.267 & 0.381 &  0.378 & 0.267 & 0.381  &  0.378 & 0.267 & 0.381 \\
 & n($\sigma_{x\%}$)  & 0.187 & 0.175 & 0.262 & 0.181 & 0.170& 0.256 & 0.216 & 0.196 & 0.288 \\
 & SMV & 0.386 & \underline{0.285} & \underline{0.405} &  0.386 & \underline{0.285} & \underline{0.405}  &  0.386 & \underline{0.285} & 0.405 \\
& NQA-QPP & \textit{0.121} & \textit{0.075} & \textit{0.115}     & \textit{0.118} & \textit{0.073} & \textit{0.109}     & 0.150 & 0.109 & 0.153     \\
& BERTQPP & 0.167 & 0.107 & 0.169     & 0.220 & 0.145 & 0.217     & 0.298 & 0.193 & 0.296     \\
& qppBERT-PL & 0.344 & 0.225 & 0.324     & 0.316 & 0.197 & 0.284     & 0.276 & 0.178 & 0.255     \\
& NQA-QPP (warm-up) & 0.187 & 0.128 & 0.186     & 0.161 & 0.107 & 0.157     & 0.287 & 0.191 & 0.282     \\
& BERTQPP (warm-up) & 0.282 & 0.187 & 0.277     & 0.234 & 0.157 & 0.233     & 0.371 & 0.251 & 0.361     \\
& qppBERT-PL (warm-up) & 0.212 & 0.151 & 0.213     & 0.167 & 0.117 & 0.170     & 0.172 & 0.115 & 0.154     \\
  \midrule
\multirow{12}{*}{CAsT-20} & Clarity & \textit{0.126}  & \textit{0.088} & \textit{0.127}  & \textit{0.126}  & \textit{0.088} & \textit{0.127} & \textit{0.126}  & \textit{0.088} & \textit{0.127} \\
 & WIG & \textbf{0.377} & \textbf{0.277} & \textbf{0.386}  & \textbf{0.377} & \textbf{0.263} & \textbf{0.373} & \underline{0.384}  & 0.264   & 0.368 \\
 & NQC & \underline{0.339} & \underline{0.261} & \underline{0.360} &  \underline{0.339} & \underline{0.261} & \underline{0.360}  &  0.339 & 0.261 & 0.360 \\
 & $\sigma_{max}$ & 0.282 & 0.219 & 0.310 & 0.282 & 0.219 & 0.310  &  0.282 & 0.219 & 0.310\\
 & n($\sigma_{x\%}$)  & 0.199 & 0.168  & 0.236  & 0.197  & 0.156 & 0.224 & 0.201 & 0.156 & 0.220 \\
 & SMV &  0.275 & 0.216  & 0.299 & 0.275 & 0.216  & 0.299  & 0.275 & 0.216  & 0.299 \\
& NQA-QPP & \textit{-0.037} & \textit{-0.037} & \textit{-0.058}     & \textit{-0.081} & \textit{-0.063} & \textit{-0.092}     & \textit{0.059} & \textit{0.023} & \textit{0.032}     \\
& BERTQPP & 0.223 & 0.157 & 0.226     & 0.216 & 0.146 & 0.212     & \textbf{0.404} & \textbf{0.281} & \textbf{0.395}     \\
& qppBERT-PL & 0.185 & 0.144 & 0.191     & \textit{0.029} & \textit{0.023} & \textit{0.031}     & 0.251 & 0.171 & 0.232     \\
& NQA-QPP (warm-up) & 0.315 & 0.218 & 0.313     & 0.240 & 0.178 & 0.245     & 0.374 & 0.267 & 0.375     \\
& BERTQPP (warm-up) & 0.253 & 0.183 & 0.257     & 0.320 & 0.236 & 0.338     & 0.349 & 0.244 & 0.346     \\
& qppBERT-PL (warm-up) & 0.218 & 0.164 & 0.227     & 0.140 & 0.115 & 0.157     & 0.348 & \underline{0.268} & \underline{0.376}     \\
  \midrule
\multirow{9}{*}{OR-QuAC} & Clarity & -0.050  & -0.029  & -0.038 & -0.050 & -0.029 & -0.038  & -0.050 & -0.029 & -0.038 \\
 & WIG & 0.137 & 0.107 & 0.145 & 0.116  & 0.088 & 0.120 & 0.140 & 0.111  & 0.149 \\
 & NQC & 0.227 & 0.163  & 0.221  &  0.227 & 0.163  & 0.221  &  0.227 & 0.163  & 0.221 \\
 & $\sigma_{max}$ &  0.442 &  0.339 & 0.443  &  0.442 &  0.339 & 0.443 &  0.442 &  0.339 & 0.443\\
 & n($\sigma_{x\%}$)  & -0.032 & \textit{-0.003} & \textit{-0.004}  & -0.073  & -0.035 &  -0.045 & \textit{-0.022} & \textit{0.008}  & \textit{0.011} \\
 & SMV & 0.098  &  0.076 & 0.106 & 0.098  &  0.076 & 0.106  &  0.098  &  0.076 & 0.106\\
& NQA-QPP & \textbf{0.615} & \textbf{0.479} & \textbf{0.615}     & \textbf{0.639} & \textbf{0.499} & \textbf{0.638}     & \textbf{0.600} & \textbf{0.470} & \textbf{0.601}     \\
& BERTQPP & \underline{0.481} & \underline{0.417} & \underline{0.541}     & \underline{0.505} & \underline{0.435} & \underline{0.563}     & \underline{0.481} & \underline{0.408} & \underline{0.529}     \\
& qppBERT-PL & 0.391 & 0.250 & 0.287     & 0.424 & 0.294 & 0.335     & 0.437 & 0.306 & 0.349     \\
 \bottomrule
\end{tabular}
\end{table*}

\begin{figure*}[!t]
  \centering
  \includegraphics[clip, trim=0cm 5.5cm 0cm 3cm,scale=0.75]{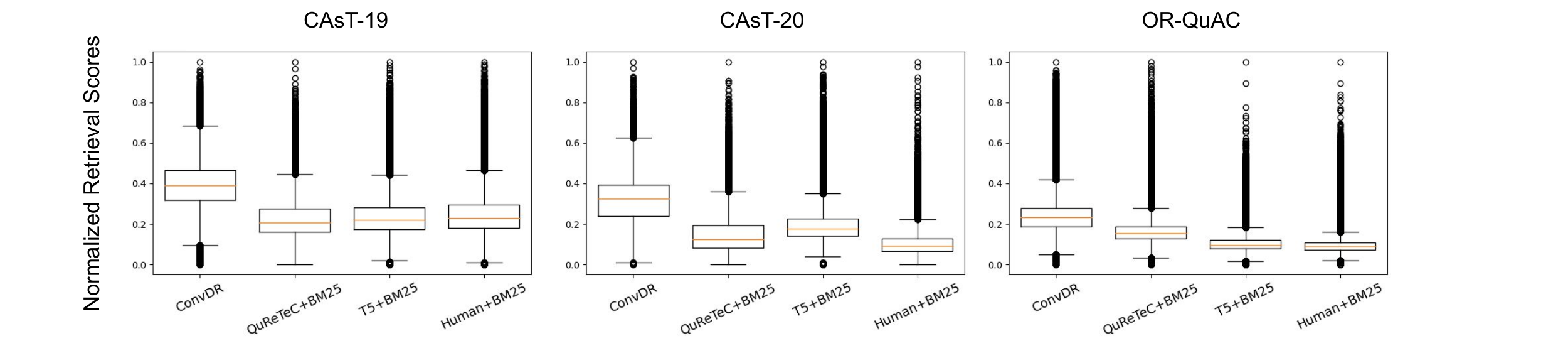}
  \caption{Distributions of retrieval scores for ConvDR and BM25 with three different rewriters on the three datasets. For the sake of comparison, we normalize the retrieval scores of a pipeline for all queries in a dataset by min-max normalization.   
}
    \label{fig:score_distribution}
\end{figure*}

First, retrieval score-based methods NQC/WIG show high effectiveness in estimating the retrieval quality of ConvDR, achieving the best performance in most cases on \CAsT-19 and \CAsT-20.
Compared to Table~\ref{tab:RQ1}, the performance of NQC/WIG is even better than their effectiveness in assessing BM25.
It contradicts the previous findings~\citep{datta2022relative,hashemi2019performance}:
\citet{datta2022relative} found that the retrieval scores from neural-based retrievers, such as ColBERT~\citep{khattab2020colbert}, are restricted within a shorter range compared to lexical-based retrievers, which may limit the performance of score-based unsupervised \QPP methods.
We speculate that there are two reasons.
First, the effectiveness of score-based methods depends on the retrieval score distribution of a specific retriever, regardless of whether they assess a lexical-based or a neural-based retriever.
Figure~\ref{fig:score_distribution} illustrates the retrieval score distributions of ConvDR and BM25 with three kinds of query rewrites in the three datasets. 
The retrieval score distribution of ConvDR displays a higher variance.
A higher standard deviation indicates that the score ranges vary more, and so the top-ranked documents are more distinguishable from the rest. 
Thus, ConvDR has a higher potential to be predicted more accurately using score-based \QPP methods. 
Second, as discussed in Section~\ref{res:rq1:overall}, score-based \QPP methods do not depend on the input queries and tend to be less impacted by the query understanding challenge in \CS.
Thus, score-based unsupervised methods show more effectiveness when assessing ConvDR compared to other supervised methods.

Second, supervised \QPP methods tend to exhibit better performance when fed with human-written query rewrites, especially on \CAsT-20, where query rewriting is much harder than \CAsT-19.
It highlights the importance of query rewriting quality.

Third, similar to our results for \ref{RQ1}, supervised \QPP methods distinctly outperform all unsupervised \QPP methods on the \ORQuAC dataset where a large amount of training data is available. 
NQA-QPP remains the state-of-the-art method on \ORQuAC.

Fourth, as with the results for \ref{RQ1}, point-wise supervised methods outperform qppBERT-PL in most cases (on \CAsT-20 and \ORQuAC).
On \CAsT-19, qppBERT-PL trained using five-fold cross-validation outperforms its point-wise counterparts warming up from \ORQuAC, showing its potential in a few-shot setting.


\subsubsection{Turn-wise \QPP effectiveness}
Similar to Section~\ref{turnbm25}, here we report the turn-wise effectiveness of
the same \QPP methods when they are fed with T5-based and human-written query rewrites to assess ConvDR.
See the two rightmost subfigures in Figure~\ref{fig:turn}.
As shown in both subfigures, the effectiveness of the score-based unsupervised methods (WIG/NQC) first exhibits lower performance at the first turn, and then shows an upward trend as conversations go on. 
In contrast, in the middle of conversations, the supervised \QPP methods are more sensitive to the actual retrieval quality; their effectiveness drops sharply as the actual retrieval quality drops.
Especially, NQA-QPP/BERT-QPP effectiveness shows a more dramatic drop from turn 4 to 6 when they are fed with T5-based query rewrites, compared to when they are fed with human-written ones.
It shows the importance of improving query rewriting quality again.
Interestingly, there is a sharp drop from turn 7 to 8 for all \QPP methods, showing the \QPP difficulty at deeper turns.






\subsection{Top ranks vs. deeper ranked lists}
To answer \ref{RQ3}, we report the results of \ref{E3} in Table~\ref{tab:RQ3}, i.e., \QPP results in terms of nDCG@3, nDCG@100, and Recall@100.
Due to space limitations, for supervised \QPP methods, we only show them in the warm-up setting.
Since qppBERT-PL works better without warm-up, we consider it both with and without a warm-up round.
We have three main observations.

\begin{table*}[]
\caption{Outcomes of Experiment~\ref{E3}.
Performance of \QPP methods on three \CS datasets: Pearson's $r$, Kendall's $\tau$, and Spearman's $\rho$ correlation coefficients with nDCG@3, nDCG@100 and Recall@100, for
estimating the retrieval quality of BM25 fed with T5-based query rewrites and ConvDR.
All coefficients are statistically significant (t-test, $p < 0.05$) except the ones in \emph{italics}.
The best value in each column is marked in \textbf{bold}, and the second best is \underline{underlined}.
}
\label{tab:RQ3}
\begin{tabular}{ll rr rr rr rr rr rr}
\toprule
 &  \multicolumn{1}{c}{}  & \multicolumn{6}{c}{T5 + BM25} & \multicolumn{6}{c}{ConvDR (\QPP fed with T5 query rewrites)} \\
  \cmidrule(lr){3-8}  \cmidrule(lr){9-14}  
 &   \multicolumn{1}{c}{}  & \multicolumn{2}{c}{nDCG@3} & \multicolumn{2}{c}{nDCG@100} & \multicolumn{2}{c}{Recall@100} & \multicolumn{2}{c}{nDCG@3} & \multicolumn{2}{c}{nDCG@100} & \multicolumn{2}{c}{Recall@100} \\
 \cmidrule(r){3-4}
 \cmidrule(r){5-6}
 \cmidrule(r){7-8}
 \cmidrule(r){9-10}
 \cmidrule(r){11-12}
 \cmidrule(r){13-14}
 & \multicolumn{1}{l}{QPP methods} & \multicolumn{1}{c}{P-$\rho$} & \multicolumn{1}{c}{K-$\tau$} & \multicolumn{1}{c}{P-$\rho$} & \multicolumn{1}{c}{K-$\tau$} & \multicolumn{1}{c}{P-$\rho$} & \multicolumn{1}{c}{K-$\tau$} & \multicolumn{1}{c}{P-$\rho$} & \multicolumn{1}{c}{K-$\tau$} & \multicolumn{1}{c}{P-$\rho$} & \multicolumn{1}{c}{K-$\tau$} & \multicolumn{1}{c}{P-$\rho$} & \multicolumn{1}{c}{K-$\tau$} \\
 \midrule
 \parbox[t]{2mm}{\multirow{10}{*}{\rotatebox[origin=c]{90}{CAsT-19}}}
& Clarity  & 0.321  & 0.234 & 0.326  & 0.257  & 0.214  & 0.187  & 0.257  & 0.176 & 0.342 & 0.227 & 0.335 & 0.216 \\
& WIG & 0.436 & 0.232 & \textbf{0.608} & \underline{0.429} & \textbf{0.579} & 0.426 & \underline{0.387} & 0.274  &  0.542 & 0.398  & 0.451 & 0.347   \\
 & NQC  & 0.348 & 0.246 & 0.548 & 0.397 & \underline{0.545} & \underline{0.444}  & \textbf{0.431} & \textbf{0.307}  & \textbf{0.647} & \textbf{0.481} & \underline{0.557}  & 0.421 \\
 & $\sigma_{max}$  & 0.442 & \underline{0.354} & \underline{0.574}  & \textbf{0.433}  & 0.494 & 0.399  &  0.378 & 0.267 & \underline{0.637} & 0.456 & \textbf{0.591} & \textbf{0.441} \\
 & n($\sigma_{x\%}$)  & 0.430 & 0.332 & 0.569 & 0.406 & 0.505  & 0.365 & 0.187 & 0.175 &  0.358 & 0.292  &0.362 & 0.288\\
 & SMV  &  0.344 & 0.250 & 0.548 & 0.417 & 0.541  & \textbf{0.466}  &  0.386 & \underline{0.285} & 0.619 & \underline{0.471}  & 0.556 & \underline{0.423} \\
 & NQA-QPP (warm-up)   & \textbf{0.538} & \textbf{0.357}   & 0.542 & 0.392   & 0.537 & 0.377   & 0.187 & 0.128   & 0.401 & 0.275   & 0.364 & 0.263 \\
 & BERTQPP (warm-up)   & \underline{0.526} & \textbf{0.357}   & 0.532 & 0.391   & 0.463 & 0.325   & 0.282 & 0.187   & 0.378 & 0.249   & 0.261 & 0.194\\
 & qppBERT-PL (warm-up)   & 0.317 & 0.218   & 0.412 & 0.279   & 0.363 & 0.263   & 0.212 & 0.151   & 0.354 & 0.233   & 0.345 & 0.249 \\
  & qppBERT-PL   & 0.414 & 0.296   & 0.509 & 0.358   & 0.452 & 0.312   & 0.344 & 0.225   & 0.461 & 0.310   & 0.455 & 0.327 \\
  \midrule
  \parbox[t]{2mm}{\multirow{10}{*}{\rotatebox[origin=c]{90}{CAsT-20}}}
& Clarity  & \underline{0.258} & 0.191  & 0.452  & 0.343  & \underline{0.467} & 0.332    & \textit{0.126} & \textit{0.088} & 0.270  & 0.195  & 0.264 & 0.178 \\
& WIG & 0.248 & \textbf{0.251} & \textbf{0.494} & \textbf{0.453} & \textbf{0.478}  & \textbf{0.438}  & \textbf{0.377}  & \textbf{0.277}  & \textbf{0.549} & \underline{0.389} & \textbf{0.465} & 0.320  \\
 & NQC  & 0.150 & \underline{0.235} & 0.363 & 0.399 & 0.320 & 0.380 & \underline{0.339} & \underline{0.261} & \underline{0.544} & \textbf{0.404} & \underline{0.463} & \textbf{0.357} \\
 & $\sigma_{max}$  & 0.179 & 0.221 &   0.339 & 0.372 & 0.339 &  0.382 & 0.282 & 0.219  & 0.496 & 0.364  & 0.440 & 0.328 \\
 & n($\sigma_{x\%}$)  &  0.178 & 0.225 & 0.413 & \underline{0.422} & 0.420 & \underline{0.410}  & 0.199 & 0.168 & 0.409 & 0.309  & 0.397 & 0.285\\
 & SMV  & 0.139 & 0.219 & 0.362 & 0.400 & 0.333 & 0.387 &  0.275 & 0.216 & 0.503 & 0.380  & 0.454 & \underline{0.352} \\
 & NQA-QPP (warm-up)  & \textbf{0.274} & 0.170   & \underline{0.471} & 0.362   & 0.466 & 0.370   & 0.315 & 0.218   & 0.310 & 0.237   & 0.324 & 0.223 \\
 & BERTQPP (warm-up)   & 0.207 & 0.171   & 0.404 & 0.301   & 0.364 & 0.246   & 0.253 & 0.183   & 0.349 & 0.242   & 0.221 & 0.133 \\
 & qppBERT-PL (warm-up)  & 0.228 & 0.213   & 0.367 & 0.305   & 0.312 & 0.287   & 0.218 & 0.164   & 0.378 & 0.272   & 0.313 & 0.229  \\
& qppBERT-PL  & \textit{0.131} & 0.125   & 0.310 & 0.251   & 0.363 & 0.275   & 0.185 & 0.144   & 0.301 & 0.217   & 0.263 & 0.196 \\
  \midrule
 \parbox[t]{2mm}{\multirow{9}{*}{\rotatebox[origin=c]{90}{OR-QuAC}}} 
& Clarity  & 0.090  & 0.085 & 0.197 & 0.196  & 0.362  & 0.312  & -0.050  & -0.029 & -0.029 & -0.015 & 0.053 & 0.057 \\
& WIG &  0.247 &  0.235& 0.376  & 0.370  & 0.482 & 0.450 &  0.137 & 0.107  & 0.195 & 0.130  &0.298 & 0.261  \\
 & NQC  & 0.251 & 0.274 & 0.356 & 0.409 & 0.414 & 0.461 &  0.227  &  0.163 & 0.302 & 0.194  & 0.402 & 0.333\\
 & $\sigma_{max}$  & 0.317 &  0.279 &  0.418 & 0.393 & 0.438 & 0.437 & 0.442 & 0.339 & 0.490 & 0.359 &0.434 & \underline{0.370} \\
 & n($\sigma_{x\%}$)  & 0.181 & 0.172 & 0.295 & 0.302 & 0.415 & 0.401  & -0.032 & \textit{-0.003} & \textit{-0.001} & \textit{0.010} & 0.102 & 0.106\\
 & SMV  & 0.204 & 0.239 & 0.311 & 0.383 & 0.396  & 0.456 &  0.098 & 0.076 & 0.170 & 0.109  & 0.313 & 0.277 \\
 & NQA-QPP   & \textbf{0.781} & \textbf{0.566}   & \textbf{0.783} & \textbf{0.602}   & \textbf{0.603} & \textbf{0.486}   & \textbf{0.615} & \textbf{0.479}   & \textbf{0.644} & \textbf{0.475}   & 0.446 & 0.323 \\
 & BERTQPP  & \underline{0.678} & \underline{0.434}   & \underline{0.767} & 0.551   & \underline{0.589} & \underline{0.484}   & \underline{0.481} & \underline{0.417}   & \underline{0.595} & \underline{0.453}   & \underline{0.447} & 0.313 \\
 & qppBERT-PL   & 0.594 & 0.507   & 0.655 & \underline{0.552}   & 0.451 & 0.440   & 0.391 & 0.250   & 0.449 & 0.277   & \textbf{0.455} & \textbf{0.383}  \\
 \bottomrule
\end{tabular}
\end{table*}

First, all \QPP methods generally perform better when predicting the retrieval quality for deeper-ranked lists. 
The estimated performance by various \QPP methods achieves a higher correlation with the actual nDCG@100/Recall@100 values in comparison with the nDCG@3 values, which is in line with~\citep{zamani2018neural}, that found predicting NDCG@20 to be harder than AP@1000.

Second, unsupervised \QPP methods get a higher correlation with nDCG@100 and Recall@100 on \CAsT-19 and \CAsT-20, showing high effectiveness in estimating the retrieval quality of deeper ranked lists.
On \ORQuAC, where training data is ample, supervised \QPP methods still keep the lead in terms of all metrics, in line with the results shown in Table~\ref{tab:RQ1} and Table~\ref{tab:RQ2}.

Third, in some cases, list-wise supervised methods outperform their point-wise counterparts when estimating the retrieval quality in terms of deeper ranked lists.
E.g., qppBERT-PL without warm-up outperforms other point-wise methods (NQA-QPP and BERTQPP with warm-up) on \CAsT-19 when assessing ConvDR in terms of nDCG@100 and Recall@100.
Also, qppBERT-PL achieves the best performance when predicting the performance of ConvDR in terms of Recall@100 on OR-QuAC.
The gains indicate that modeling a list of retrieved items has the potential of benefiting the retrieval quality estimation for deeper-ranked lists.


\section{Related work}

\textbf{Query performance prediction.}
The \acf{QPP} task is to estimate the retrieval quality of a search system in response to a user query without relevance judgments \cite{he2006query,carmel2010estimating}. 
\QPP methods have shown a high correlation with the retrieval quality in the context of ad-hoc retrieval. 
They can help to obtain better-performing retrieval pipelines in different ways, including query routing \cite{sarnikar2014query}. 
Moreover, query difficulty signals have been used to provide direct feedback to users, allowing them to reformulate queries or seek alternative information sources if the results are expected to be poor. 
%

Typically, \QPP methods can be classified into pre- and post-retrieval methods~\citep{carmel2010estimating}. 
Pre-retrieval methods estimate query performance based on the query and corpus statistics before retrieval takes place. 
Post-retrieval methods use additional information from the ranked list to predict query performance after retrieval. 
In this paper, we focus on post-retrieval \QPP methods because they have shown superior performance compared to pre-retrieval methods in most cases.
Post-retrieval \QPP methods include both supervised and unsupervised methods.

Traditional \QPP methods have mostly relied on an unsupervised approach where query term frequency and corpus statistics are used as indicators for query performance \cite{shtok2010using,shtok2012predicting,he2006query,he-2008-using,hauff2010query,zhou2007query,hauff2008survey}.
More recent studies model \QPP by deep learning-based models. 
These studies have shown that supervised methods for \QPP are more effective than unsupervised \QPP approaches in an ad-hoc retrieval setting. 
These supervised methods require a significant amount of data and training instances, such as the MS MARCO dataset \cite{nguyen2016ms}, to perform \QPP effectively~\cite{hashemi2019performance,arabzadeh2021bert,zamani2018neural,datta2022pointwise}.
To the best of our knowledge, \QPP has mostly been limited to ad-hoc retrieval tasks. 
\citet{hashemi2019performance} explore the ability of \QPP methods to predict performance for non-factoid question answering. 
Studies of the performance of \QPP methods in \CS, have been limited. 


\header{Conversational search}
\Acf{CS} is the task of retrieving relevant passages in response to user queries in a multi-turn conversation~\citep{dalton2020cast}.
A unique challenge in \CS is that a user query in a conversation is context-dependent, i.e., it may contain omissions, coreferences, or ambiguities, making it challenging for ad-hoc search methods to capture the underlying information need~\citep{radlinski2017theoretical}. Recovering the underlying information need from the conversational history is crucial~\citep{mao2022curriculum}.
To address this challenge, there are two main groups of \CS methods, namely, \textit{query-rewriting-based retrieval} and \textit{conversational dense retrieval}.
Query-rewriting-based retrieval methods first rewrite a query that is part of a conversation into a self-contained query and then feed it to an ad-hoc retriever~\citep{mele2020topic,voskarides2020query,yu2020few,vakulenko2021comparison,lin2021multi,wu2022conqrr}.
Query rewriting can be conducted by either term expansion or sequence generation.
The former adds terms from the conversational history to the current query, e.g., by designing rules~\citep{mele2020topic} or training a binary term classifier~\citep{vakulenko2021comparison}, while the latter directly generates the reformulated queries using pre-trained generative language models, e.g., GPT-2~\citep{yu2020few} and T5~\citep{lin2021multi}.
 
Conversational dense retrieval methods train a query encoder to encode the current query and the conversational history into a contextualized query embedding; the contextualized query embedding is expected to implicitly represent the information need of the current query in a latent  space~\citep{qu2020open,yu2021few,lin2021contextualized,mao2022curriculum,kim2022saving,mao2022convtrans}.
\citet{lin2021contextualized} train the query encoder by optimizing a ranking loss over a large number of pseudo-relevance judgments. 
\citet{yu2021few} train the query encoder to mimic the embeddings of human-written queries output by the query encoder of the ad-hoc dense retriever ANCE~\citep{xiongapproximate}.
\citet{mao2022curriculum} train the query encoder to denoise noisy turns in the conversation history by contrastive learning.

Little research has been done into \QPP for \CS.
\citet{roitman2019study,arabzadeh2022unsupervised} explore \QPP in single-turn \CS, where they use \QPP to help a \CS system take the next appropriate action given a user query.
Specifically, they use \QPP to assess the retrieved answer quality to determine whether the system should return the answer to the user.
\citet{lin2021multi,al2022improving} use \QPP to improve the retrieval quality of a \CS system.
\citet{lin2021multi} use a \QPP method to determine whether the current query should be expanded with keywords from the previous turns. \citet{al2022improving} use \QPP methods to select the better query rewrite from different ones.
\citet{meng2023Performance} investigate the performance of pre-retrieval \QPP methods when they estimate the retrieval quality of BM25 fed with T5-generated query rewrites.
Also, \citet{meng2023Performance} propose to incorporate query rewriting quality to improve \QPP effectiveness.
Additionally, \citet{vlachou2022performance} explore \QPP in the context of conversational fashion recommendation, which differs from \CS.

What we add to the studies listed above, is a comprehensive reproducibility study where we reproduce various \QPP methods designed for ad-hoc search systems in the setting of multi-turn \CS.

\section{Conclusion}
In this reproducibility study, we examined whether three key findings for \QPP in ad-hoc search hold in \CS.
We experimented with \QPP methods designed for ad-hoc search in three \acs{CS} settings:
\begin{enumerate*}[label=(\roman*)]
\item predicting the retrieval quality of BM25 while studying the impact of query rewriting; 
\item predicting the retrieval quality of a conversational dense retrieval method, namely ConvDR; and
\item predicting the retrieval quality for top ranks vs.~ deeper-ranked lists.
\end{enumerate*} 

We found that the three findings on \QPP for ad-hoc search do not generalize to \CS very well.
Specifically, we found
\begin{enumerate*}[label=(\roman*)]
\item supervised \QPP methods distinctly outperform their unsupervised counterparts only when a large amount of training data is available, while unsupervised \QPP methods show strong performance when being in a few-shot setting and predicting the retrieval quality for deeper ranked lists;
\item point-wise supervised \QPP methods outperform their list-wise counterparts in most cases; however, list-wise \ac{QPP} methods are more data-efficient, show a slight advantage in predicting the retrieval quality for deeper ranked lists; and 
\item retrieval score-based unsupervised \QPP methods show high effectiveness in estimating the retrieval quality of a conversational dense retrieval method, ConvDR, either for top ranks or deeper ranked lists;
the effectiveness of score-based methods relies on the retrieval score distribution of a specific retriever, regardless of whether they assess a lexical-based or a neural-based retriever.
\end{enumerate*} 

Our paper reveals that feeding T5 or QuReTeC query rewrites into \QPP methods to estimate the retrieval quality of \CS methods exhibits great performance.
We also identify the drawbacks of \QPP methods designed for ad-hoc search in the context of \CS, motivating the next direction for the modeling of \QPP for \CS. 
We show that the quality of query rewriting is of great importance, highlighting the need to improve query writing quality. 
It also shows the need to develop a mechanism of conversational context understanding for \QPP methods to directly understand raw historical utterances.
Also, we reveal that the data sparsity problem in \CS severely reduces the performance of supervised \QPP methods.
Thus, designing \QPP methods using few-shot learning techniques is one possible way.

We point to two limitations of our study, namely, 
\begin{enumerate*}[label=(\roman*)]
\item we only consider estimating the retrieval quality of one conversational dense retrieval method, and 
\item we only use correlation metrics to evaluate the performance of \QPP methods.
\end{enumerate*} 
In future work, we plan to
\begin{enumerate*}[label=(\roman*)]
\item consider more conversational dense retrieval methods such as CQE~\citep{lin2021contextualized} as well as other dense retrieval methods for \CS, such as T5-based rewriter+ANCE~\citep{xiongapproximate}, and 
\item introduce \QPP-specific evaluation metrics, such as scaled Absolute Ranking Error (sARE) and scaled Mean Absolute Ranking Error (sMARE)~\citep{faggioli2022smare,faggioli2021enhanced}.
\end{enumerate*}



\begin{acks}
We would like to thank our reviewers for their feedback.
This research was partially supported by the China Scholarship Council (CSC) under grant number 202106220041, and the Hybrid Intelligence Center, a 10-year program funded by the Dutch Ministry of Education, Culture and Science through the Netherlands Organisation for Scientific Research, \url{https://hybrid-intelligence-centre.nl}.

All content represents the opinion of the authors, which is not necessarily shared or endorsed by their respective employers and/or sponsors.
\end{acks}

\clearpage
\bibliographystyle{ACM-Reference-Format}
\balance
\bibliography{References}


\end{document}